# Carrier-envelope-phase-independent field sampling of single-cycle transients using Homochromatic Attosecond Streaking

## H. Y. Kim, Z. Pi, and E. Goulielmakis*

*Institute of Physics, University of Rostock, Albert-Einstein-Str. 23, 18059 Rostock, Germany*
*Corresponding author: e.goulielmakis@uni-rostock.de*

**The recent development of Homochromatic Attosecond Streaking (HAS) has enabled a novel, highly precise method for ultrafast metrology of attosecond electron pulses as well as for real-time sampling of the instantaneous field waveforms of light transients. Here, we evaluate the potential of HAS as a method for precisely sampling the field waveform of non-phase-stabilized, single-cycle transients of light. We show that the extreme nonlinearity of field emission allows our technique to isolate and track the waveform of a single carrier-envelope phase (CEP) setting whose field dynamics results in the most energetic electron cutoff. Our results establish HAS as a robust, compact, all-solid-state method for characterizing light fields with attosecond-level precision and as a powerful tool in light field synthesis.**

The complete temporal characterization of optical fields has been central in the study of strong-field phenomena and the advancement of attosecond science [1]. This has motivated the development of numerous field-sensitive characterization methods over the past two decades [2–8]. Among these, extreme ultraviolet (EUV) attosecond streaking stands out, as it provides direct access into the instantaneous "absolute field strength" of a light transient with attosecond precision [2–4]. A significant step towards a compact, all solid-state, absolute-field-sensitive characterization of light transients was recently taken with the development of Homochromatic Attosecond Streaking (HAS) [9]. Since HAS does not require the generation of extreme ultraviolet attosecond pulses for sampling the field waveform and does not rely on gas phase measurements it presents a strong potential for the realization of ultracompact field characterization devices. Here we show a new important feature of HAS namely that it can access the waveform of pulses whose carrier envelope phase is not stabilized.

To show how this is possible, it is worth briefly revisiting the physical principles of HAS (Fig. 1(a)). An intense laser pulse (the pump, red curve in Fig. 1(a)) releases an electron burst from the apex of a metal nanotip via field-driven tunneling. As is typically described in strong field recollision models [10–15], the liberated electron is driven by the pump laser pulse to backscatter off the tip surface at a recollision time. Given that the recollision occurs within a half-cycle of the pump field, the attosecond confinement of the back-scattered wavepacket can serve as a tool to trace an optical waveform as discussed later on in this text. Upon recollision with the tip surface the electron wavepacket gains further momentum in the laser field reaching the detector with a final momentum. A sufficiently weak replica of the same pulse (the gate, red dashed curve in Fig. 1(a)) the intensity of which is approximately two orders of magnitude lower than that of the pump pulse, and thus cannot ionize the tip, modifies the momentum and thus energy of the strong field-emitted electronic wavepacket [9,16,17] resulting in a shift of its final momentum (blue dashed curves in Fig. 1(a)). We have previously shown that the final momentum shift is proportional to the effective vector potential $A_{HAS}(t)$ of the gate pulse acting on the wavepacket [9]. This vector potential is explicitly associated with the vector potential of the gate field $A_g(t)$,

$$A_{HAS}(t) = 2A_g(t) - \frac{1}{\Delta t}\int_{-\Delta t}^{0} A_g(t+t')dt' \quad (1)$$

where $\Delta t$ is the excursion time from the ionization to recollision moment of the electron. An explicit relationship between $A_{HAS}(t)$ and $A_g(t)$ can be best expressed in Fourier domain, $\tilde{A}_{HAS}(\omega) = \tilde{A}_g(\omega)\tilde{g}(\omega)$ where the multiplier $\tilde{g}(\omega)$ is given by $\tilde{g}(\omega) = \left[2 - \frac{1}{\omega\Delta t}(e^{-i\omega\Delta t}-1)\right]$ (2).

For a sufficiently weak gate, the cut-off energy variation on a HAS spectrogram can be approximated as being linearly proportional to the HAS vector potential $A_{HAS}(t)$, analogous to conventional EUV attosecond streaking. The relationship of Eq. 1 allows to evaluate the gate field $A_g(t)$ from the HAS vector potential $A_{HAS}(t)$ and thus the electric field waveform of the desired pulse. The right panel of Fig. 1(a) shows semiclassical simulations of a HAS spectrogram for a recorded single-cycle pulse waveform highlighting the shift of the cutoff energy of the electrons spectra depending on the delay between pump and gate fields. The effective vector potential $A_{HAS}(t)$ shown as a blue dashed curve is accessed by tracking the modulation of the modified cutoff energy as a function of the delay.

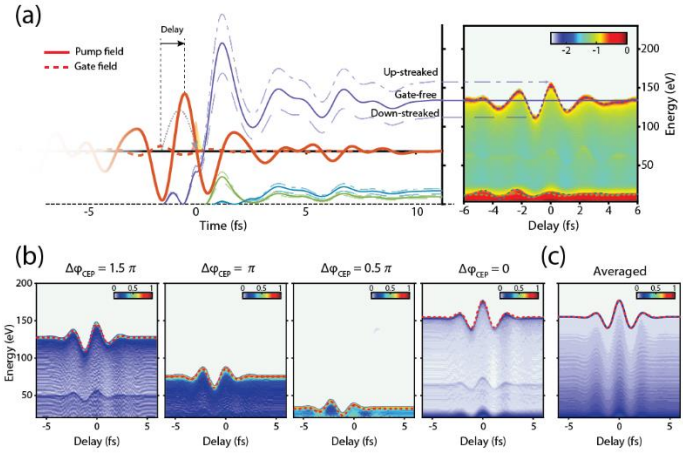

Fig. 1. (a) Illustration of Homochromatic Attosecond Streaking (HAS) principle. Simulated HAS spectrograms for different carrier-envelope phases of a single-cycle optical driving pulse (b) and an incoherent average of 40 spectrograms over CEP settings from 0 to 2π (c).

For HAS to be able to track the field waveform of laser pulses whose carrier-envelope phase (CEP) is unstabilized from pulse to pulse, it is critical that the electron cutoff energy, over which $A_{HAS}(t)$ is evaluated in a HAS trace, stems from or is dominated by a single CEP setting of the pulse that field-ionizes the tip. However, a sequence of non-phase-stabilized pulses will yield a HAS spectrogram comprised of an incoherent superposition of spectrograms associated with each individual CEP phase of pulses that field-ionize the nanotip. To better illustrate why this is possible, Fig. 1(b) shows HAS spectrograms simulated for four representative settings of the carrier-envelope phase of the same pulse, where $\Delta\varphi_{CEP}$ denotes relative phase from the CEP setting for the highest cutoff energy. The evaluated $A_{HAS}(t)$ vector potential is indicated by the red dashed curve in each phase setting. A careful comparison of these spectrograms reveals significant differences in spectral intensity (as indicated by the false color), the overall structure of electron cutoffs, and the highest cutoff energies.

Fig. 1(c) shows a spectrogram formed by the incoherent average of the HAS spectrograms in Fig. 1 (b) including several more (a total of 40) intermediate phase settings to increase fidelity. A comparison of the individual spectrograms and the averaged one suggests that the phase setting $\Delta\varphi_{CEP} = 0$ in Fig. 1(b), dominates the response. This becomes evident when directly comparing the vector potential evaluated from the average HAS spectrogram (blue curve in Fig. 1(c)) with that of the most energetic cutoff (red dashed curve), which shows remarkable agreement.

To explore the above ideas experimentally we implement HAS according to the schematic of Fig. 2(a). We used linearly polarized single-cycle optical pulses (central energy, $\hbar\omega_L \approx 1.7$ eV) composed in a light field synthesizer [4,18,19]. The laser pulse is reflected off a dual concave mirror assembly which consists of concentric nickel coated, inner and outer mirrors as shown in Fig. 2(a) inset. This allows the division of the beam of the pulse into an intense inner beam (pump field) and weak outer beam (gate field), who are in turn focused onto an electrically grounded, tungsten nanotip (apex radius of ∼ 35 nm). A piezo translation stage on which the inner mirror is attached allows to control the delay between pump and gate pulses with attosecond precision. We compose HAS spectrograms by recording electron spectra as function of the delay between pump and gate pulses using a time-of-flight spectrometer whose entrance is located ∼ 3 mm above the tungsten nanotip. A representative HAS spectrogram, recorded for the phase setting of the synthesized waveform that yielded the highest cutoff energy (above 100 eV), is shown in Fig. 2(b). White and blue curves indicate $A_{HAS}(t)$ and the evaluated electric field respectively according to using the Eq. 1 and 2.

Fig. 2(c)-(e) show HAS traces and corresponding field waveforms recorded for several more CEP settings of the driving pulse under identical conditions. High fidelity in synthesis and control of the generated waveforms is suggested by the comparison of red dashed and blue solid curves in each panel, representing the predicted shape of the waveform (based on the field evaluated in Fig. 2(b)) and the one measured upon change of the CEP setting of the pulse. The high degree of fidelity of both synthesis and measurement is more formally verified by evaluating the degree of similarity (D.S.) among measured and numerically predicted field waveforms [20]. Indeed, a degree of similarity > 0.87 (1 corresponds to identical

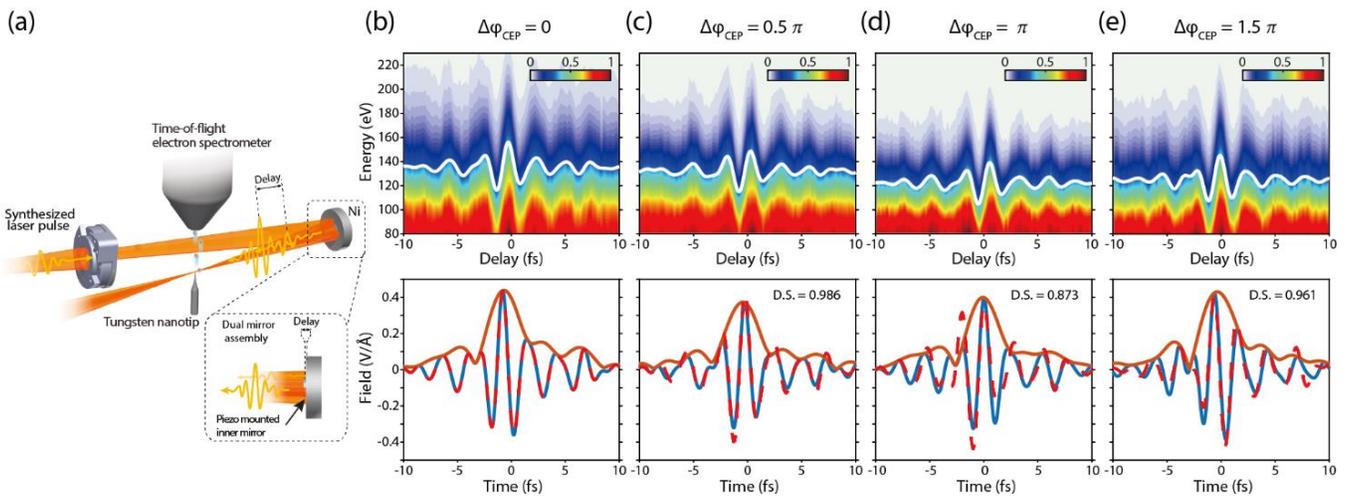

Fig. 2. (a) schematic of experimental setup for Homochromatic Attosecond Streaking (HAS). HAS spectrograms recorded at various CEP settings of the driving pulse, $\Delta\varphi_{CEP} = 0$ (b), 0.5π (c), π (d) and 1.5π (e) (upper panels). The white curves indicate the cut-off energy modulation in the HAS spectrogram along the time delay, which is associated with the HAS vector potential $A_{HAS}(t)$. (lower panels) Gate fields retrieved from the measured HAS vector potential.

waveforms and 0 to uncorrelated waveforms) is evaluated. More importantly variations both spectral intensity as well as of the cutoff energy among different CEP settings are clear evident in reasonable agreement with the semiclassical predictions of Fig. 1(b).

In the next step we turned the CEP stabilization of the driving laser pulse off, and we recorded a HAS spectrogram as shown in Fig. 3(a). The spectrogram shows a well discernible pulse structure, highlighted by the evaluated $A_{HAS}(t)$ that does not show significant qualitative changes with the measurements taken under fixed CEPs. To verify that the trace of Fig. 3(a) probes the single CEP setting that yields the highest electron cutoff we compared the evaluated field waveform of Fig. 3(a) with that retrieved from spectrogram of Fig. 2(b) as well as from the average of spectrograms recorded at CE phases over 0 to 2π (Fig. 3(b)). The results of Fig. 3(c) show a stunning similarity of ∼ 0.98 between waveforms retrieved in the cases of non-stabilized CEP (blue curve) and a fixed CEP ($\Delta\varphi_{CEP} = 0$) for the highest cut-off energy, (yellow dashed curve).

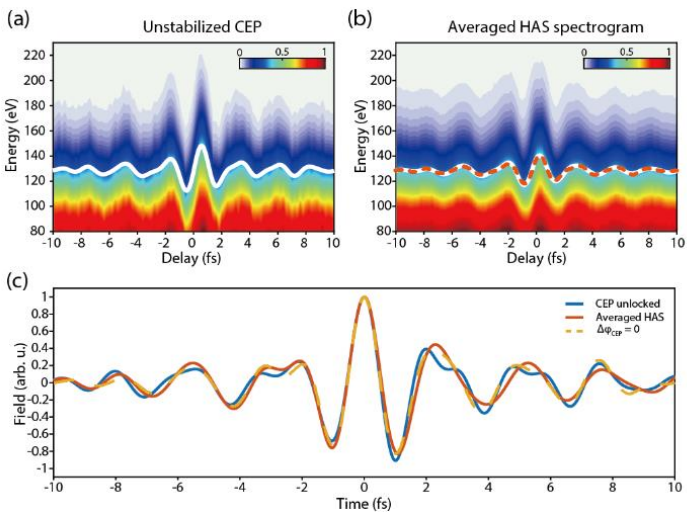

Fig. 3. (a) HAS spectrogram recorded with non-stabilized pulse-to-pulse CEP. (b) averaged spectrogram of HAS spectrograms recorded at eight different CEP settings from 0 to 2π with 0.25π steps. (c) Comparison of field waveforms retrieved from spectrograms in the cases of CEP non-stabilized, averaged and a stabilized CEP at $\Delta\varphi_{CEP} = 0$.

In this letter, we demonstrated that HAS can provide a reliable method for complete characterization of optical fields including pulses with random CEP. This possibility can become essential for laser systems of very low repetition rate whose phase stabilization is challenging. More importantly HAS also in the case of unlocked CEP could allow direct study of dynamics of attosecond phenomena as those manifested directly in the field waveform of light [21].

**Funding.** Funded by the Deutsche Forschungsgemeinschaft (DFG, German Research Foundation) - SFB 1477 "Light-Matter Interactions at Interfaces" (project number 441234705) and by the European Research Council (ERC) under the European Union's Horizon Europe research and innovation programme (Advanced Grant Agreement No. [Project 101098243].

**Disclosures.** The authors declare no conflicts of interest.

**Data availability.** Data underlying the results presented in this paper are not publicly available at this time but may be obtained from the authors upon reasonable request.